\documentclass[12pt]{article}%
\usepackage{amsmath}
\usepackage{amsfonts}
\usepackage{amssymb}
\usepackage{graphicx}%
\newtheorem{theorem}{Theorem}
\newtheorem{corollary}{Corollary}

\begin{document}

\author{Renan Cabrera, Herschel Rabitz \\ 
{\small Department of Chemistry, Princeton University, Princeton, New Jersey 08544, USA}}

\title{
The landscape of quantum transitions driven by single-qubit unitary 
transformations with implications for entanglement}

\author{Renan Cabrera, Herschel Rabitz\\
\small Department of Chemistry, Princeton University, Princeton, New Jersey 08544, USA \\
\small rcabrera@princeton.edu
}

\maketitle 
%\abstract{
\begin{abstract}

 This paper considers the control landscape of quantum transitions
in multi-qubit systems driven by unitary transformations with single-qubit 
interaction terms. The two-qubit case is fully analyzed to reveal the features of the landscape 
including the nature of the absolute maximum and minimum, the saddle points,
 and the absence of traps. The results permit calculating the Schmidt state starting from an 
arbitrary two-qubit state following the local gradient flow. The analysis of multi-qubit 
systems is more challenging, but the generalized Schmidt states also may be located by 
following the local gradient flow. Finally, we show the relation between the generalized 
Schmidt states and the entanglement measure based on the Bures distance.

\end{abstract}
%\pacs{03.67.Mn, 03.65.Ud, 02.30.Yy}

\maketitle 

\emph{\small This is an author-created, un-copyedited version of an article accepted for publication in J. Phys. A: Math. Theor. IOP Publishing Ltd is not responsible for any errors or omissions in this version of the manuscript or any version derived from it. The definitive publisher authenticated version is available online at 10.1088/1751-8113/42/27/275303}

\section{Introduction}
The topology of quantum control landscapes is important because it establishes 
the general features of the control behavior generated by applying 
external fields \cite{PhysRevA.37.4950,chakrabarti2007qcl}. 
The landscape for quantum transitions, assuming complete 
controllability, was analysed with the conclusion that there are no traps
\cite{rabitz2004qoc,rabitz2004cqp,rabitz2006ocl,rabitz2006toc} that could
hinder achieving the highest possible control outcome.
This paper studies the problem of describing the landscape of quantum 
transitions driven by local unitary operators, i.e., those acting 
on one qubit at a time, for multi-qubit systems \cite{schulteherbrueggen2008gfo}. 

The Schmidt states, defined for pure bi-partite systems, are important
because of the insight they can provide about entanglement.  
The Schmidt states were generalized in 
\cite{PhysRevLett.85.1560,carteret:7932}, in order to treat multipartite
systems. This paper will show how to obtain the canonical form of the 
generalized Schmidt states by following the local gradient flow. This technique 
ultimately leads to a method to measure the entanglement of pure systems 
based on the optimal implementation of local unitary operations as a 
subset of the more general classical operations and classical communication 
protocols as it was pursued with other methods 
\cite{PhysRevLett.83.436,mandilara:022331}.  

It is convenient to define the following bracket operation
\begin{eqnarray}
  \langle X  \rangle_0  &=&  \frac{1}{2N} Tr[ X + X^\dagger ] 
\end{eqnarray}
The comparative fidelity between two density matrices, when at least one 
of them is pure is  $\langle \rho_0 \rho_T  \rangle_0$. If one of the states is driven by a unitary 
operator, then the  cost function  can be written as
\begin{equation}
 F =  \langle U^\dagger \rho_0 U \rho_T  \rangle_0,
\end{equation}
This expression has  the same form as the cost function for the optimization 
of the expectation value of an observable $\mathcal{O}$ \cite{rabitz2006ocl},
\begin{equation}
  \mathcal{J}_1 =  \langle U\rho_0 U^\dagger \mathcal{O}  \rangle_0,
\end{equation}
which was the subject of prior landscape studies \cite{chakrabarti2007qcl}. 
The fidelity function for the state transfer can be rewritten as
\begin{equation}
 F =  \langle  \rho_0 U \rho_T U^\dagger  \rangle_0.
\end{equation}
An infinitesimal transformation of the unitary operator can be expressed as
\begin{equation}
 U \rightarrow U^\prime = U e^{\delta A} = U(1+\delta A),
\end{equation}
with $\delta A$ being an anti-Hermitian element that lies in the corresponding Lie algebra, 
so that an infinitesimal variation of $U$ becomes
\begin{equation}
 \delta U = U \delta A,
\end{equation} 
which can be used to calculate the first order variation of the 
fidelity as
\begin{equation}
  \delta F = \langle \rho_0 U[ \delta A , \rho_T]U^\dagger  \rangle_0. 
\end{equation}
A subsequent manipulation results in
\begin{equation}
 \delta F =  \langle [\rho_T, U^\dagger \rho_0 U ] \delta A  \rangle_0
 =  \langle  [\rho_T, U^\dagger \rho_0 U ]    U^\dagger \delta U  \rangle_0
\end{equation}
thereby identifying the gradient as
\begin{equation}
Grad_1 = U [ U^\dagger \rho_0 U , \rho_T ], 
\label{gradient-1}
\end{equation}
with the corresponding gradient flow equation
\begin{equation}
\frac{dU}{ds} =   U [ U^\dagger \rho_0 U , \rho_T ].
\label{gradient-flow-1}
\end{equation}

The fidelity can be expanded up to second order to obtain
the quadratic form for the Hessian
\begin{equation}
  \delta^2F = \langle \{\rho_T, U^\dagger \rho_0 U \} (\delta A)^2 \rangle_0
  - 2 \langle U^\dagger \rho_0 U  \delta A \rho_T \delta A  \rangle_0, 
\end{equation}
where $\{,\}$ stands for the anti-commutator. This quadratic form
is simplified at the critical points where the gradient (\ref{gradient-flow-1})
is zero
\begin{equation}
  \delta^2F|_{c} = 2( \langle \rho_T U^\dagger \rho_0 U (\delta A)^2 \rangle_0
  - \langle U^\dagger \rho_0 U  \delta A \rho_T \delta A  \rangle_0). 
\end{equation}

The local gradient flow is found by eliminating multi-qubit terms in
$\delta A$, such as $\sigma_3\otimes \sigma_3$ and leaving single qubit terms, such 
as $\sigma_3 \otimes \mathbf{1}_{2 \times 2}$ or $\mathbf{1}_{2\times2} \otimes \sigma_3 $. In this way, only strictly localized interactions are 
involved as happens in classical mechanics. Defining $\mathcal{P}$ as the projector that 
eliminates multi-qubit terms, the variation of the unitary operator 
with the corresponding local flow is
\begin{equation}
 \delta U = U \mathcal{P}  \delta A.
\end{equation}
The projector $\mathcal{P}$ is easily calculated by tracing
one-qubit terms. For example, the two-qubit projector is 
\begin{equation} 
\mathcal{P} =  \frac{1}{4} \sum_{j=1}^{3}
  Tr[ \, \cdot \,\, \sigma_0 \otimes \sigma_j]\sigma_0 \otimes \sigma_j+
  Tr[ \, \cdot \,\, \sigma_j \otimes \sigma_0]\sigma_j \otimes \sigma_0, 
\label{projector-su2su2}
\end{equation}
with $\sigma_0 = \mathbf{1}_{2\times 2}$, so that $\mathcal{P} \delta A $ is constrained to
 the six-dimensional Lie algebra $su(2)\times su(2) \subset  su(4)$. 
The first order variation subject to the local flow becomes
\begin{equation}
 \delta F =  \langle  [\rho_T, U^\dagger \rho_0 U ] \mathcal{P}  U^\dagger \delta U  \rangle_0
 =  \langle  \mathcal{P} \left( [\rho_T , U^\dagger \rho_0 U] \right)  U^\dagger \delta U  \rangle_0,
\end{equation}
which results in the following local gradient 
\begin{equation}
Grad_{1}^{local} = U \mathcal{P} [ U^\dagger \rho_0 U , \rho_T].
\label{local-gradient-1} 
\end{equation}

\section{ Two-Qubit Systems}
The Schmidt states play an important role in the quantification of the 
entanglement of two-qubit systems. We will show their importance
in describing the quantum landscape characterized by the local gradient flow
and then calculate the Schmidt state of a given entangled state by following
the local gradient flow (excepting the maximally entangled state).

Consider the landscape where the target state is a Schmidt state denoted as
 $\rho_T = \rho_S(\theta)$. The Schmidt states for two-qubit systems can be parametrized with a single 
variable as
\begin{equation}
 | \psi_{\rho_S}  \rangle = \cos (\theta/2) | \uparrow \uparrow  \,\rangle 
  + \sin (\theta/2) | \downarrow \downarrow  \,\rangle, 
\end{equation}
whose corresponding density matrix reads
\begin{equation}
  \rho_S (\theta) = 
  \begin{pmatrix} \cos^2( \theta/2 )     & 0 &  0 &  \frac{1}{2}\sin \theta  \\
 0 & 0 & 0 & 0 \\
 0 & 0 & 0 & 0 \\
          \frac{1}{2}  \sin \theta  & 0 & 0 & \sin ^2( \theta/2 )   \end{pmatrix},
\end{equation}
with  $0 \le \theta \le \pi $, in the standard basis $\{
  | \uparrow    \uparrow   \, \rangle , 
  | \uparrow    \downarrow \, \rangle ,
  | \downarrow  \uparrow   \, \rangle  ,
  | \downarrow  \downarrow \, \rangle \}$.

The critical states $\rho_c = U_c^\dagger \rho_0 U_c $ obey the following equation
\begin{equation}
  \mathcal{P} [ \rho_c , \rho_S(\theta)] =0
 \label{critical-condition} 
\end{equation}
It can be shown that this equation is satisfied by critical states 
that fall into one of the following two cases 
\begin{itemize}

\item Another Schmidt state $\rho_c=\rho_S(\phi)$. In this case, the eigenvalues of the 
Hessian around the critical points are either negative or mixed, with 
the following explicit form
\begin{equation}
h(\theta,\phi) = 
\begin{pmatrix} 0 \\
    -1 - \cos(\theta - \phi) - \sin\theta - \sin\phi\\
    -1 - \cos(\theta - \phi) - \sin\theta - \sin\phi\\
    -4 \sin\theta \sin \phi\\
   -1 - \cos(\theta - \phi) + \sin\theta + \sin\phi\\
 -1 - \cos(\theta - \phi) + \sin\theta + \sin\phi
\end{pmatrix},
\end{equation}
For each critical state with a negative 
spectrum $h(\theta_0, \phi_0)$, there is another one with a mixed spectrum  $h(\theta_0, \pi-\phi_0)$. 
Conversely, for each critical state with a mixed spectrum  $h(\theta_0, \phi_0)$, there is another 
one with a negative spectrum  $h(\theta_0, \pi-\phi_0)$.
So, for each initial state there is a pair of critical states that can be reached 
by following the local gradient flow,
such that one of them is a saddle point and the other is an stable maximal point. 
If the initial state is separable, the two possible 
critical states are given by $\rho_S(0)$ or $\rho_S(\pi)$.  

\item The critical sub-manifold spanned by the basis $\{ | \uparrow \downarrow  \,\rangle ,
  | \downarrow  \uparrow \,\rangle \}$ with the
following explicit form of the critical state  
\begin{equation}
 \rho_c = x | \downarrow  \uparrow \,\rangle  \langle \downarrow  \uparrow | + 
 (1-x)  | \uparrow  \downarrow \,\rangle  \langle \uparrow  \downarrow |, 
\end{equation}
where the eigenvalues of the Hessian are
\begin{equation}
 \begin{pmatrix}
  1- \sqrt{(1-2x)^2\cos^2 \theta+\sin^2\theta}\\ 
 1- \sqrt{(1-2x)^2\cos^2 \theta+\sin^2\theta}\\
 0\\
 1+ \sqrt{(1-2x)^2\cos^2 \theta+\sin^2\theta}\\
 1+ \sqrt{(1-2x)^2\cos^2 \theta+\sin^2\theta}\\
0
 \end{pmatrix},
\end{equation}
which corresponds to a positive spectrum, associated with
the minimum.

\end{itemize}

Based on the features of the critical points we can state the 
following theorem
\begin{theorem}
The fidelity landscape between a pure separable state $\rho_0$
and a target Schmidt state $\rho_S(\theta)$ (with $\theta \neq \pi/2$) has saddle points
but no traps. Moreover, the separable states that maximize 
the fidelity converge to  either  $ | \uparrow \uparrow  \,\rangle $ or  $ | \downarrow \downarrow  \,\rangle  $ depending on the target 
state as they follow the local gradient flow, according to the 
following formula   
\begin{equation}
 \lim_{U \rightarrow U_c}{U^\dagger \rho_0 U  } = 
 \begin{cases}  | \uparrow \uparrow  \,\rangle \langle \uparrow \uparrow  | &  0 < \theta < \pi/2 \\
                 | \downarrow \downarrow  \,\rangle \langle \downarrow \downarrow  | &  \pi/2 < \theta < \pi
    \end{cases},
\label{critical-states}
\end{equation}
\end{theorem}
This theorem is a direct result of the fact that those limiting states
are the only Schmidt states with zero entanglement. Moreover, we can also say
that 
\begin{corollary}
For pure states, the maximum fidelity between an entangled state and a
separable state can be calculated from the corresponding Schmidt state
$ | \psi_{S}  \rangle = \cos (\theta/2) | \uparrow \uparrow  \,\rangle 
  + \sin (\theta/2) | \downarrow \downarrow  \,\rangle$ as
\begin{equation}
 \mathcal{F}(\theta) = \max F =
  \begin{cases} \cos^2( \theta/2) & \theta \le \pi/2 \\
        \sin^2( \theta/2) & \pi/2 <\theta \le \pi 
  \end{cases}
\label{optimal-fidelity}
\end{equation}
\end{corollary}

\begin{figure} 
\centering
\includegraphics[scale=0.5]{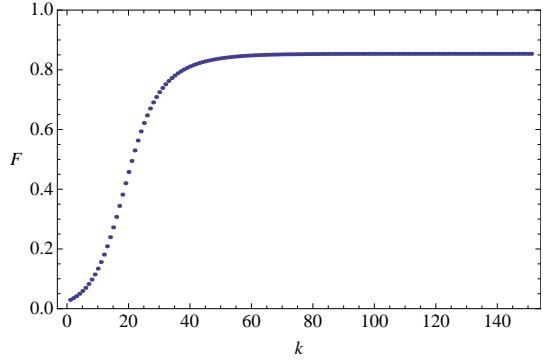}
\caption{ Fidelity curve of the states following the local gradient flow for
the initial separable state (\ref{arb-sep-state}) with $\rho_S(\pi/4)$ as the target state. 
The fidelity never reaches 1 but attains the global maximum associated with the 
limiting state $ | \uparrow \uparrow  \,\rangle $.  }
\label{fig:fidelity-aproaching-sep}
\end{figure}

\begin{figure} 
\centering
\includegraphics[scale=0.6]{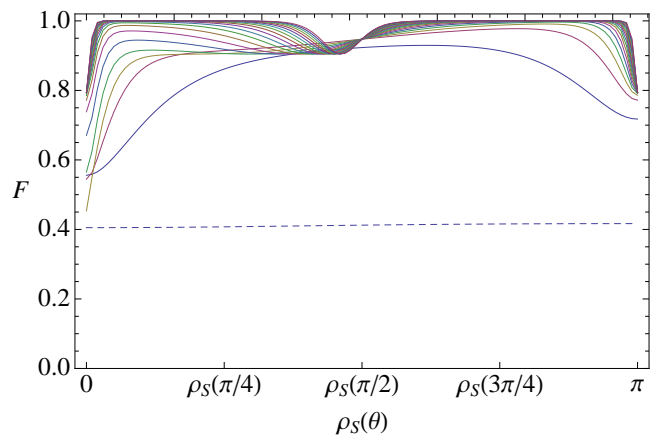}
\caption{ Fidelity of a random entangled state moving towards its Schmidt state 
as a function of the target Schmidt state $\rho_S(\theta)$ employed to drive the local gradient flow. 
The dashed line represents the fidelity of the initial random state with respect to its
Schmidt state and each subsequent curve corresponds to another step in the approach 
by following the local gradient flow. The figure suggests 
that the arbitrary state never reaches its corresponding Schmidt state when the
gradient employs the target states  $\rho_S(0)$, $\rho_S(\pi/2)$  and  $\rho_S(\pi)$. }
\label{fig:fidelity-convergence}
\end{figure}

The maximum fidelity $\mathcal{F}(\theta)$ can be used to calculate the Bures 
distance as the entanglement measure, which satisfies all the features required for
a good entanglement monotone \cite{vedral1997qe,PhysRevA.57.1619}. 
In the present case of pure two-qubit systems the entanglement formula is
\begin{equation}
E_B(\rho) = 
2\left( 1 - \sqrt{ \mathcal{F}(\theta} \right)
\label{bures-entanglement}.
\end{equation}
As a first example, Figure \ref{fig:fidelity-aproaching-sep} shows the fidelity
of the states following the local gradient flow for the initial separable 
state described by
\begin{equation}
 \rho_0 = e^{\frac{i}{4\pi} \sigma_0 \otimes \sigma_1} | \uparrow \uparrow  \,\rangle \langle \uparrow \uparrow  | 
  e^{-\frac{i}{4\pi} \sigma_0 \otimes \sigma_1}
\label{arb-sep-state}
\end{equation}
with  $\rho_S(\pi/4)$ as the target state and $ | \uparrow \uparrow  \,\rangle $ as the limiting state. 
The next example considers the following entangled initial state
\begin{equation}
 \rho_0 = 
 e^{\frac{i}{\pi/4} \sigma_2 \otimes \sigma_0}
  e^{\frac{7i}{ 10 \pi} \sigma_2 \otimes \sigma_2} | \uparrow \uparrow  \,\rangle \langle \uparrow \uparrow  | 
   e^{-\frac{7i}{ 10 \pi} \sigma_2 \otimes \sigma_2}
 e^{-\frac{i}{\pi/4} \sigma_2 \otimes \sigma_0}
\label{entan-state}
\end{equation}
driven by the local unitary flow with $\rho_S(\pi/4)$ as the target state,
and the following limiting Schmidt state 
\begin{equation}
\lim_{U \rightarrow U_c}{ U^\dagger \rho_0 U} =
\begin {pmatrix}
                    0.793893 & 0 & 0 & 0.404508 \\
                    0 & 0 & 0 & 0 \\
              0 & 0 & 0 & 0 \\
        0.404508 & 0 & 0 & 0.206107
  \end{pmatrix}.
\end{equation}
Almost any Schmidt state can be used as the target state in order to drive the 
local gradient flow, excepting those with 
 $\theta=\{0,\pi/2,\pi\}$, because of convergence issues. For example, Figure 
\ref{fig:fidelity-convergence} shows how the arbitrary state (\ref{entan-state}) 
 approaches its Schmidt state for the range of  target Schmidt states.

The local gradient flow was driven by employing target Schmidt states, 
but the landscape is invariant under the
application of local unitary operations on both the initial 
and target state. The local unitary transformations include
local phases, that are able to change the phase of the
Schmidt states. This means that the
general stable critical states are Schmidt states with
the possibility of extra phases. For example, consider the following 
arbitrary entangled state made from a Schmidt state and local 
unitary transformations 
\begin{equation}
 \rho_E = e^{\frac{i\pi}{4}\sigma_2 \otimes \sigma_0} 
  e^{\frac{i\pi}{4}\sigma_0 \otimes \sigma_1} \rho_S(\pi/4)
  e^{-\frac{i\pi}{4}\sigma_0 \otimes \sigma_1}
  e^{-\frac{i\pi}{4}\sigma_2 \otimes \sigma_0}. 
\end{equation}
The initial separable state is taken as  $\rho_i = | \uparrow \uparrow  \,\rangle \langle \uparrow \uparrow  | $. The local gradient flow 
converges to a separable unitary operator $U_c$ with the following corresponding
separable state
\begin{equation}
 \rho_c = U_c^\dagger \rho_i U_c =\begin{pmatrix} 1/2&-1/2\\-1/2&1/2 \end{pmatrix} \otimes
          \begin{pmatrix} 1/2&-i/2\\i/2&1/2\end{pmatrix}.
\end{equation}
This state can be diagonalized by the following local unitary operator
\begin{equation}
 T =  \begin{pmatrix} 1/\sqrt{2}& 1/\sqrt{2}\\-1/\sqrt{2}&1/\sqrt{2} \end{pmatrix} \otimes
          \begin{pmatrix} i/\sqrt{2}&i/\sqrt{2}\\-1/\sqrt{2}&1/\sqrt{2}\end{pmatrix},
\end{equation}
such that $T^\dagger \rho_c T =  | \uparrow \uparrow  \,\rangle \langle \uparrow \uparrow  | $. This suggests that $T$ could
be used to reduce $\rho_E$ to its expected Schmidt state $\rho_S(\pi/4)$, but instead 
we obtain a Schmidt state with an extra phase $-i$
\begin{equation}
T^\dagger \rho_E T = \begin {pmatrix} 
                     \cos ^2\left (\frac {\pi } {8} \right) & 0 & 0 & -i\cos \left (\frac {\pi } {8} 
                       \right) \sin \left (\frac {\pi } {8} \right) \\
                     0 & 0 & 0 & 0 \\
               0 & 0 & 0 & 0 \\
          i\cos \left (\frac {\pi } {8} \right) \sin \left (\frac {\pi } {8} 
    \right) & 0 & 0 & \sin ^2\left (\frac {\pi } {8} \right)
  \end {pmatrix}.
\end{equation}
This extra phase can be eliminated by the use of local phase transformations, which 
otherwise leave the absolute value of the components of the density matrix invariant.

\section{Three or More Qubit Systems}
The entanglement in a two-qubit system can be minimally characterized by 
a single variable as shown in the Schmidt state. The number of variables needed 
to parametrize a n-qubit system is $2^{n+1}-2$ up to a global phase, and the number 
of variables to parametrize a single qubit is $3n$, thus, the minimum number 
of variables needed to parametrize the entanglement of an n-qubit system is
\begin{equation}
N_E = 2^{n+1}-2-3n,
\end{equation} 
which is five for three-qubit systems. 
The canonical form of the generalized Schmidt states is important because of 
the information that can be obtained about entanglement 
\cite{PhysRevA.65.052302,acin2001tqp,pan2005eet,PhysRevA.65.052302}.
A canonical form of the generalized Schmidt state for three-qubits 
was introduced in \cite{PhysRevLett.85.1560} as
\begin{equation}
 |\psi_S \rangle = 
 \lambda_1 | \uparrow \uparrow \uparrow  \rangle + 
 \lambda_2 e^{i\phi} | \uparrow \downarrow \downarrow  \rangle +
 \lambda_3 | \downarrow \uparrow \downarrow  \rangle +
 \lambda_4 | \downarrow \downarrow \uparrow  \rangle+
 \lambda_5 | \downarrow \downarrow \downarrow  \rangle,
\label{Schmidt-threeQ}
\end{equation}
with $\lambda_i \ge 0 $, $\phi \ge 0$ and $\sum |\lambda_i|^2 = 1$. The canonical form of the generalized 
Schmidt state for n-qubit systems was given in \cite{carteret:7932} indicating 
that the missing basis elements in the generalized Schmidt state are 
\begin{equation}
  | \downarrow \uparrow \uparrow ... \uparrow  \rangle,  
  | \uparrow \downarrow \uparrow ... \uparrow  \rangle,
  | \uparrow \uparrow \downarrow ... \uparrow  \rangle,
  ...
  | \uparrow \uparrow \uparrow ... \downarrow  \rangle.
\label{missing-Schmidt}
\end{equation}

The landscape of multi-qubit systems is richer and more complex than the
two-qubit case. Considering the case where the initial state
is separable and following the reasoning in \cite{carteret:7932}, 
we can always demand that $\lambda_1 \ge \lambda_k$. However, the analysis is simpler
if we relax some generality and demand that $\lambda_1>\lambda_k$, for $k>1$.
The variation of the fidelity can be written as
\begin{equation}
 \delta F = \delta \langle \Psi | \psi_S  \rangle \langle \psi_S | \Psi  \rangle
   = 2 Re[  \langle \Psi | \psi_S   \rangle \langle \psi_S| \delta| \Psi  \rangle  ] 
\label{variation-Psi}
\end{equation}
The canonical form in (\ref{Schmidt-threeQ}) indicates that if we start with 
a generic separable state $ | \Psi \rangle =| \psi_1  \rangle \otimes | \psi_2  \rangle \otimes | \psi_3  \rangle $ and allow local unitary 
transformations, the isolated maximum fidelity is achieved at
the critical state $ | \Psi_c \rangle = | \uparrow \uparrow \uparrow  \,\rangle $. The first order variation under 
local unitary transformations is made of a linear combination of 
basis elements with at most one qubit reversed,
\begin{equation}
  \delta | \uparrow \uparrow \uparrow  \rangle  =  
(1 + i\delta_1 ) | \uparrow \uparrow \uparrow  \rangle +  
    \delta_2 | \downarrow \uparrow \uparrow \rangle + 
    \delta_3 | \uparrow \downarrow \uparrow \rangle +
     \delta_5 | \uparrow \uparrow \downarrow \rangle, 
\end{equation}
with $\delta_1 \in \mathbb{R}$, $\delta_2 \in \mathbb{C}$, $\delta_3 \in \mathbb{C}$, $\delta_5 \in \mathbb{C}$. We can use this variation 
in order to evaluate $\delta F$ given by (\ref{variation-Psi}) at the critical
state  $ | \Psi_c \rangle $ and verify that it is a stationary point,
thus justifying the canonical form of the generalized Schmidt state. 
The missing basis elements (\ref{missing-Schmidt}) 
form a critical sub-manifold associated with the fidelity minimum of 
zero value. The generic identification of the remaining critical states 
is difficult and depends on the specific $\lambda_j$ values. However, 
if $\lambda_j>0$, then there are no additional critical
states because the  aforementioned critical states exhaust all the
possibilities to obtain $\delta F=0$. 

As a concrete example, consider calculating the generalized 
Schmidt state of the following arbitrary state
\begin{equation}
 |\psi_T \rangle = \begin{pmatrix}
0.3+0.1i \\ 0.2\\0.3\\0.3\\0.4\\0.2\\0.5\\ \sqrt{1-0.77} 
\end{pmatrix}.
\end{equation}
Following the same procedure used in the two-qubit case,
 we  use the local gradient flow to calculate the optimized separable 
state $|\psi_c \rangle$ that maximizes the fidelity $|\langle \psi_T|\psi_c \rangle|^2$,
 starting form an initial separable state (e.g. $| \uparrow \uparrow \uparrow  \,\rangle$). 
The optimized state $|\psi_c\rangle$ can be diagonalized using a local unitary 
transformation. Applying the same local unitary transformation to the
target state $|\psi_T \rangle \langle \psi_T |$ we  obtain
\begin{equation}
  | \hat{\psi}_S \rangle = 
\begin{pmatrix}
 0.986657 \\
 0\\
 0\\
 -0.125609-0.0245643 i\\
 0\\
 0.0151643- 0.0312796 i\\
0.0703562+ 0.0477398 i\\
-0.0138602 + 0.0387071 i
\end{pmatrix}, 
\end{equation}
which is almost in the canonical form (\ref{Schmidt-threeQ}). 
The first component can always be put in real form by choosing a suitable
global phase. The remaining procedure is to employ the three available
local phase transformations in order to eliminate the phase of 
last three components to finally obtain
 \begin{equation}
  | \psi_S \rangle = 
\begin{pmatrix}
 0.986657 \\
 0\\
 0\\
 -0.125609-0.0245643 i\\
 0\\
0.0347616\\
0.085024\\
0.0411138
\end{pmatrix},
\end{equation}
which we ascertain to be the global maximum because $|\lambda_1|$ is greater
than the rest of the components. The local phase transformations do 
not change the absolute value of the components of the column spinor, 
so, it is easy to verify that, for example, in the 
last component $|-0.0138602 + 0.0387071 i| =0.0411138 $. 

The procedure to calculate the Schmidt state can be used to calculate the
Bures distance as an entanglement measure if $|\lambda_1|$ is  greater than
the rest of the components. In this case the formula of the Bures distance 
as a measure of entanglement is simply
\begin{equation}
 E_B(\rho) = 2( 1 - |\lambda_1|  ).
\end{equation}
The study of higher multi-qubit states follows along the same general lines 
of the three-qubit state. Thus, we are able to calculate 
the generalized Schmidt state as well as the Bures distance as a measure
of entanglement for most of the cases where $\lambda_1$ results in a value
greater than the rest of the components.

\section{Conclusions}
The landscape of local quantum transitions for two-qubit systems is
well suited for optimization through the gradient flow because 
of the lack of traps. We showed how to extend these results 
to muli-qubit systems and presented an example on how to calculate the
generalized Schmidt state for three-qubits. 
The local gradient flow can be easily applied to higher multi-qubit systems 
and even though we could not give a complete analysis of the landscape,
a criteria was presented to establish if the global maximum was attained.
A generalization of this analysis to mixed multi-qubits is desireable, 
but this is a much more challenging problem because of the severe 
limitations that unitary transformations present.

\section*{Acknowledgments}
The authors acknowledge support from the DOE.

\section*{References}

\bibliographystyle{unsrt}
\bibliography{SchmidtDistance}

\end{document}